\documentclass[aps,prl,twocolumn,superscriptaddress,nofootinbib]{revtex4-1}
\usepackage{amssymb, graphicx, hyperref}
\usepackage[intlimits]{amsmath}
\usepackage[english]{babel}
\begin{document}

\title{On the accuracy of using Fokker Planck equation in heavy ion collision}

\author{Nirupam Dutta}
\affiliation{Theoretical Physics Division, Variable Energy Cyclotron Centre,
1/AF, Bidhan Nagar, 
Kolkata 700~064, India}
\email[E-mail:]{nirupamdu@gmail.com}
\author{Trambak Bhattacharyya}
\affiliation{Visiting Researcher, Indian Institute of Technology, Indore, Madhya Pradesh-452017, India}

\date{\today}

\begin{abstract}

Application of Fokker-Planck equation to heavy quark transport in the evolving medium created in heavy ion collision is critically scrutinised. We realise that the approach introduces a moderate uncertainty in drag and diffusion coefficients culminating in huge ambiguity in the theoretical prediction of nuclear modification factor $R_{AA}$. Quantitative estimation of the error is presented by considering recent developments in this field.

\end{abstract}

\maketitle

The last thirty years have witnessed fervent activity in the physics community aiming to understand de-confined quark matter created through high energy nucleus-nucleus collisions. Ideas started to percolate when studies of finite temperature quantum chromodynamics (QCD) \cite{itoh,perry1975,parisi1975} suggested the existence of a phase where the quark and gluonic degrees of freedom appear instead of being confined within hadrons. Another advertisement that has been inviting further attention to heavy ion collisions (HIC) is the scope to study our early Universe from the convenience of a laboratory. But, one should be cautious because in early Universe the evolution of quark gluon plasma (QGP) was quite slow compared to that in HIC \cite{iancu,hatsuda,gelis}. The time scale of evolution of QGP in early Universe is expected to be of the order of micro-seconds \cite{hatsuda} whereas the QGP in a laboratory does not even persist for more than $10-12$fm \cite{gelis}. It is, therefore, a matter of debate how far one can infer about the early Universe from a study of quark gluon plasma in a laboratory where it is eventually subjected to a violent evolution. But this ambitious attempt may wait! First and foremost, we need to determine how well such a fast evolving medium can be understood with the tools at our disposal.

Clear and convincing evidence for the formation of a strongly interacting medium was announced by experiments in 2010 \cite{atlas2010}. Theoretical and experimental projects have been on run to figure out various properties of this medium. To explore such a short-lived medium, one has to rely on certain internal probes . Ones like heavy quarks are born much before the creation of QGP and witness the whole of its evolution. Essentially, the heavy quarks lose energy inside the medium and this phenomenon of energy loss, the jet quenching, is a well established signal for the formation of such a medium \cite{bjorken}. Obviously, this energy loss relies on the transport of heavy quarks through the medium. With the help of transport coefficients, the change in initial momentum distribution, $f_{in}$, of heavy quarks can be computed by using Boltzmann transport equation or Fokker-Planck equation (FPE). Hence, the relative change in distribution $\frac{f_{f}}{f_{in}} $,  $f_{f}$ designating the final distribution of heavy quarks, yields the nuclear suppression factor, $R_{AA}$, an experimentally measurable quantity.

The FPE can be represented in the following form:
\begin{equation}
\frac{\partial f}{\partial t}= \frac{\partial}{\partial p_i}\left[A_i(\vec{p})f+\frac{\partial}{\partial p_j}[B_{ij}(\vec{p})f]\right]\label{fpe},
\end{equation}
where $\vec p$ characterises momentum of the heavy quark with a momentum distribution $f$ . $A$ and $B$ are the drag and diffusion coefficients respectively at a fixed temperature of the medium. Although, in heavy ion collision it is still not beyond suspicion whether the medium gets enough time to thermalise, yet to make life easier one usually advocates for a local thermal equilibrium. The application of FPE to a static QGP is very straight forward for momentum independent transport coefficients (drag and diffusion) \cite{vankampen} and the solution to eq.\ref{fpe} has been deduced a number of times in this context. Some studies have been done by considering momentum dependent drag and diffusion as well \cite{surasree}.

The scenario in heavy ion collision is drastically different from that of a fixed temperature thermal bath because the temperature of QGP falls off rapidly as the fireball expands. A clever way of incorporating this into the Fokker Planck equation has been adopted in recent literature\cite{svetitsky,surasree,rapp2005,santosh,munshi,greco} for both momentum independent and momentum dependent transport coefficients. There, the drag $A$ and diffusion $B$ in eq.\ref{fpe} are replaced by time dependent functions owing to their implicit dependence on time through the temperature. This scheme tacitly introduces a time scale $\Delta t$ within which the variation in temperature is so small that Fokker Planck equation for a fixed temperature bath can be safely applied. In principle, $\Delta t$ must tend to zero to justify the replacement of $A$, $B$ by $A(t)$, $B(t)$. But, it is not permissible as the FPE is only valid at a time scale $\tau$ which should be much bigger than the collision time $\tau_{c}^{HQ}$ 
\footnote{$\tau_{c}^{HQ}$ is the time for which the heavy quark comes into the ' sphere of influence' of the bath particle}
and much smaller than the relaxation time $\tau_{r}^{HQ}$ of heavy quarks in medium. Hence $\Delta t$ should at least be of the order of $\tau$. Furthermore, to guarantee collisions, $\Delta t$ should be considered not less than the heavy quark mean free time $\tau_{f}^{HQ}$. So, we need to check whether these considerations are compatible with the present scenario of rapidly evolving QGP in heavy ion collisions.

In centre of momentum (COM) frame, $\tau_{c}^{HQ}$ can not be less than the inverse of average transverse momentum transferred per collision. We go with the usual convention by considering a gluon dominated plasma and hence, the averaging may be performed with the help of the heavy quark (Q)-gluon(g) differential scattering 
cross-section,

\begin{equation}
\frac{d \sigma}{d\hat{t}}=\frac{2\pi\alpha^2}{\hat{t}^2}(1-\Delta_M^2)^2 \label{diffcross}
\end{equation}
where $\alpha$ is the strong coupling. For a heavy quark with mass $M$, COM collision energy $\sqrt{\hat s}$, $\Delta_M^2$= $M^2/\hat{s}$. $\hat{s}$ and $\hat{t}$ are the Mandelstam variables.

Using eq. \ref{diffcross}, the square of the average transverse momentum transferred is derived to be,
\begin{equation}
\langle q_{\bot}^{2}\rangle = \frac{\hat s m_{D}^{2}}{\hat s-4m_{D}^{2}}ln\frac{\hat s}{4 m_{D}^{2}}~. \label{qperp}
\end{equation}
Here $m_{D} = g T \sqrt{\left( C_{A}+\frac{N_{f}}{2}\right) \frac{1}{6}}$ with $g=\sqrt{4\pi\alpha}$ is the Debye mass at temperature $T$. $C_{A}=3$ is the Casimir factor and $N_{f}=2$ is the number of quark flavours contributing to the gluon self energy. The collision time $\tau_{c}^{HQ}$ being of the order of $\langle q_{\bot}^{2}\rangle^{-\frac{1}{2}}$ varies from $0.3 - 0.2$ fm in the temperature range $200-400$ MeV of QGP. Collision time can also be given by $\frac{1}{gT}$ for soft gluon exchange processes \cite{beraudo}, in which case it lies between $0.5-0.25$fm for the same temperature variation. To ensure the collision of heavy quarks with medium particles, $\tau(\Delta t)$ should be taken larger than the mean free time $\tau_{f}^{Q}$ of heavy quark, the latter lying between $1.5-0.99$fm. The relaxation time of heavy quarks $\tau_{r}^{HQ}$ is the inverse of the product of the gluon density $n_{g}(=16\zeta(3)T^3/\pi^2)$ and scattering cross-section $\sigma$, \cite{thoma,biro}. Its value is around $1.45-0.72$fm, given again the same temperature range.


The last paragraph contains details of the relevant time scales pertaining to heavy quarks. Another crucial time scale is the mean free time $\tau_{f}^{Med}$ of medium or bath particles which can be evaluated similarly from the gluon-gluon scattering cross section and thereby ranges from $1-0.5$fm for the selected  temperature variation. To describe heavy quark transport using FPE, local thermal equilibrium of QGP is necessary and to assure it the observational time scale $\tau$ should be much larger than $\tau_{f}^{Med}$. Therefore, the following inequality has to be satisfied,
\begin{equation}
\tau_{r}^{HQ}\gg\tau\gg\tau_{f}^{Med}~.
\end{equation}
Only if these time scales are sufficiently well separated, the medium effects will be manifested through the transport coefficients. On the contrary, we see that the time scales here are of the same order. Thus we are compelled to ask whether it is possible to choose an observation time scale $\tau$ within the petite window between $\tau_{f}^{Med}$ and $\tau_{r}^{HQ}$. This casts a doubt on the applicability of FPE even to a static QGP. Moreover, for an evolving medium, one has to ensure temporal local thermal equilibrium. For the time being, let us ignore this subtle issue. There is a far more serious problem with the accuracy of transport equation when applied to evolving QGP as is illustrated in the remaining of this Letter.

 Even if the medium is taken to be in temporally local thermal equilibrium, the time scale of FPE has to be much larger than $0.2$fm for QGP. There also remains the restriction (mentioned earlier) that the change in temperature must not be appreciable during $\Delta t$, whose lower limit is set by $\tau$. This concept was implicitly used in earlier articles \cite{surasree,santosh,rapp2005,rapp2013}. Let us take $\tau = 0.5$fm which is midway between $\tau_{c}^{HQ}$ and $\tau_{r}^{HQ}$ . The corresponding change in temperature in heavy ion collision is about $15-20$MeV \cite{heinz}. If we calculate the resulting uncertainty in time dependent drag and diffusion coefficients ($\Delta A,~\Delta B$) around their respective mean values within the temperature range of $200-400$MeV then as per some literature the uncertainty will be not less than $6-10$\% \cite{santosh,tb} while others estimate the same to be more than $10-15$\% \cite{svetitsky,rapp2005} for each. The proportional error ($\Delta R_{AA}$) in theoretical estimation of $R_{AA}$ due to the uncertainty in drag and diffusion values can be evaluated from the final momentum distribution $f_{f}(p)$. For an initial momentum distribution centred around $p_{0}$, the final distribution is,
\begin{equation}
f_{f}(p,t)=\left[ \frac{A}{2 \pi B} \left(1-e^{-2At}\right) \right]^ {\frac{1}{2}} \exp \left[\frac{A}{2B}\frac{(p-p_{0}e^{-At})^2}{1-e^{-2At}} \right]\label{1d}.
\end{equation}
The above is a solution to the one dimensional FPE \cite{svetitsky}. The proportional error that $R_{AA}$ acquires is 
\begin{equation}
\Delta R_{AA}= \frac{\Delta A}{A}+\frac{1}{2}\frac{\Delta B}{B}+\epsilon_{1}^{p,p_{0}}(\Delta A, \Delta B)+\epsilon_{2}(\Delta A, \Delta B,t).
\end{equation}
Here, $\epsilon_{1}$ contributes to the error through its dependence on initial and final momenta of heavy quarks whereas $\epsilon_{2}$ contains error accrued due to time dependence. The first two terms account for the uncertainty irrespective of the momenta of heavy quarks and time.
 Consequently, we find the minimum percentage error or uncertainty in the prediction of $R_{AA}$ to be more than $10-18$\% depending on different temperature dependences of transport coefficients quoted in recent literature\footnote{We have quoted those articles which together cover the entire range of uncertainty. The error calculated from unquoted literature lies within the specified range. } \cite{svetitsky,rapp2005,santosh,tb}. Furthermore, this error propagates in time. Thus we see that the accumulated error in nuclear modification factor for an evolving QGP persisting for $10-12$fm can be huge enough to question the accuracy of FPE.

To sum things up, we realise that FPE can be applied to a slowly evolving medium if its time scale of evolution is much larger than $\tau$. In that case, the temperature change is negligible within $\Delta t$ and hence, the transport coefficients remain unchanged. Then, no question of uncertainty in $R_{AA}$ arises and we can happily employ the clever technique of introducing time dependent functions $A(t)$, $B(t)$. But in heavy ion collisions, we are faced with the reverse scenario as the medium evolves very rapidly. The moderate uncertainty gained by the transport coefficients and the ensuing considerable uncertainty in nuclear suppression factor $R_{AA}$ bear testimony to this fact. It remains to ascertain whether the accuracy to which such transport equations predict heavy quark energy loss is satisfactory relative to ongoing experiments with far improved precision. Probably no!

Our argument makes it clear that to mitigate the incompatibility of various time scales, the simplest option is to think of a description where the time scale $\tau$ can be made arbitrarily small (so that change in temperature within that interval is negligible). One such approach has been recently devised in the context of heavy quark bound states (bottomonium) \cite{borghini} by modelling them as open quantum systems. Development along this line for open heavy quarks is deferred for our future investigation.


\vspace{2cm}
\begin{acknowledgements}
We are thankful to Nicolas Borghini for his critical reading of the manuscript. We acknowledge Surasree Mazumder and Md. Younus for discussions. T. B. acknowledges support from the DST sponsored project No. SR/MF/PS-01/2014- IITI(G) entitled "A Large Ion Collider Experiment (ALICE) upgrade, operation and utilization"for support.
\end{acknowledgements}
  

\end{document}